\documentclass[conference]{IEEEtran}
\IEEEoverridecommandlockouts
\usepackage{cite}
\usepackage{amsmath,amssymb,amsfonts}
\usepackage{algorithmic}
\usepackage{graphicx}
\usepackage{textcomp}
\usepackage{xcolor}
\usepackage{xspace}
\usepackage{multirow}
\usepackage{array}
\usepackage{booktabs}
\usepackage{tikz}

\usepackage{subcaption}
\usepackage{wrapfig}
\usepackage{placeins} 

\def\BibTeX{{\rm B\kern-.05em{\sc i\kern-.025em b}\kern-.08em
    T\kern-.1667em\lower.7ex\hbox{E}\kern-.125emX}}

\newcommand{\design}{SpANNS\xspace}

\newcommand{\tikzcircle}[1]{\tikz[baseline=(char.base)]{
            \node[shape=circle,draw,fill=black,inner sep=1pt] (char) {\textcolor{white}{#1}};}}
\newcommand{\tikzcirclewhite}[1]{\tikz[baseline=(char.base)]{
            \node[shape=circle,draw,inner sep=1pt] (char) {#1};}}

\begin{document}

\title{SpANNS: Optimizing Approximate Nearest Neighbor Search for Sparse Vectors Using Near Memory Processing
}



\author{
Tianqi Zhang, Flavio Ponzina, Tajana Rosing \\
University of California, San Diego \\
\{tiz014, fponzina, tajana\}@ucsd.edu
}

\maketitle
\begin{abstract}
Approximate Nearest Neighbor Search (ANNS) is a fundamental operation in vector databases, enabling efficient similarity search in high-dimensional spaces. While dense ANNS has been optimized using specialized hardware accelerators, sparse ANNS remains limited by CPU-based implementations, hindering scalability. This limitation is increasingly critical as hybrid retrieval systems—combining sparse and dense embeddings—become standard in Information Retrieval (IR) pipelines.
We propose \design, a near-memory processing architecture for sparse ANNS. \design combines a hybrid inverted index with efficient query management and runtime optimizations. The architecture is built on a CXL Type-2 near-memory platform, where a specialized controller manages query parsing and cluster filtering, while compute-enabled DIMMs perform index traversal and distance computations close to the data. It achieves 15.2$\times$ to 21.6$\times$ faster execution over the state-of-the-art CPU baselines, offering scalable and efficient solutions for sparse vector search. 
\end{abstract}


\section{Introduction}

Nearest Neighbor Search (NNS) is a fundamental problem in high-dimensional data processing, involving the identification of the closest points to a given query. It is widely applied across various domains, including recommendation systems, multimedia databases, and document retrieval. To meet the demands of large-scale datasets and real-world applications with strict latency requirements, Approximate Nearest Neighbor Search (ANNS) techniques have been developed, balancing efficiency and accuracy for similarity searches.

\begin{figure}[tp] 
\centering
\includegraphics[width=\linewidth]{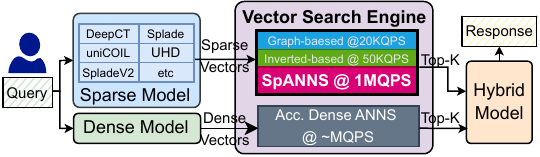} 
\vspace{-0.3cm}
\caption{Sparse Vector ANNS in the Vector Search Engine for Information Retrieval Pipelines.}
\vspace{-0.6cm}

\label{fig:intro}
\end{figure}

A key application of ANNS is in vector search engines—core components of modern Information Retrieval (IR) pipelines. These systems typically index embeddings to support fast similarity search. Prior work has focused heavily on dense ANNS, driven by the success of dense embeddings in capturing rich semantic content~\cite{ir2}. This has led to optimized dense-vector accelerators and index structures widely adopted in production systems. However, there is a growing trend toward hybrid retrieval models~\cite{hybrid_m1, ir1,ir2} that combine dense and sparse embeddings, motivated by the complementary strengths of each and build the pipeline as shown in Figure~\ref{fig:intro}. Sparse embeddings—such as those produced by SPLADE~\cite{splade, spladev2}, DeepCT~\cite{deepct}, and uniCOIL~\cite{unicoil}—are more interpretable, efficient, and effective at capturing rare or out-of-domain terms~\cite{uhd}. These hybrid systems require both dense and sparse ANNS backends, yet most existing ANNS systems and accelerators remain optimized only for the dense case.

\begin{figure*}[tp] 
\centering
\includegraphics[width=\textwidth]{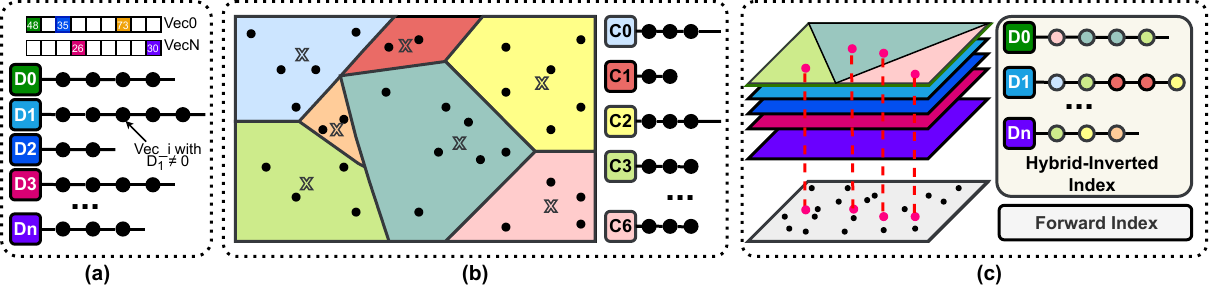} 
\vspace{-0.4cm}
\caption{Illustration of Inverted Index Approaches: (a) Content-Based, (b) Clustering-Based, and (c) Hybrid Design.}
\vspace{-0.3cm}
\label{fig:ivf}
\end{figure*}

Within this pipeline, the vector search engine is critical, tasked with efficiently searching over massive datasets. As shown in Figure~\ref{fig:intro}, dense vector ANNS methods~\cite{diskann,hnsw,scann,faiss} have been the focus of significant research and development and are widely integrated into vector database (DB) systems~\cite{pg, milvus}. Supported by specialized accelerators like GPUs~\cite{gpu_graph, faiss}, FPGAs~\cite{fpga_anns}, near-storage\cite{proxima, ndsearch}, and near-memory technologies~\cite{graph_nmp,cxl-anns}, dense vector search engines achieve throughput in the range of millions of queries per second (QPS). 
In contrast, sparse vector search engines remain underexplored, often relying on CPU-based implementations with only tens of thousands of QPS. While sparse matrix multiplications (SpMM) have been highly optimized for accelerators~\cite{arch_spmm}, they are inefficient for sparse vector ANNS, requiring comparisons against the entire reference set~\cite{spmm_search}. Similar to dense vector ANNS, inverted index~\cite{wand,seismic,bridge} and graph-based methods~\cite{pg,pyanns} have been explored for sparse embeddings. Inverted index-based methods, such as Weighted AND (WAND)~\cite{wand} and BWM~\cite{BWM}, selectively skip irrelevant portions of the dataset, while sketching methods~\cite{sketch} approximate inner product distances to enhance computational efficiency. 

However, those operations are severely limited by the memory system and the lack of specialized hardware support, leaving a significant gap in throughput compared to dense ANNS accelerators. 
As shown in Figure~\ref{fig:intro}, dense accelerators~\cite{graph_nmp} achieve millions of QPS, while sparse graph~\cite{pyanns} and inverter-based search~\cite{knowhere,seismic} engines lag in throughput. This gap persists despite the computational advantages of sparse embeddings, particularly in large-scale applications where interpretability and efficiency are critical.

Our work introduces \design, a hardware-software co-designed accelerator that bridges this gap:
\begin{itemize}
    \item We design \design, the first accelerator for inverted index-based sparse vector NNS, which is 15$\times$ to 21$\times$ faster than a Seismic CPU baseline~\cite{seismic}. To the best of our knowledge, no previous work has developed such an accelerator for sparse vector DB search engines to date.
    \item We develop a near memory processing (NMP) hybrid inverted index that combines non-zero dimensions and clustering-based indexing for efficient sparse ANNS.
    \item We design a CXL Type-2 NMP architecture with a controller for query management and cluster filtering. The specialized DIMMs support efficient index processing and runtime optimization, retaining their functionality as normal memory.
\end{itemize}

\section{Related Works: Inverted index-based ANNS}
The inverted index was originally designed to speed up search queries by quickly identifying and ranking documents containing relevant terms.  Formally, for each node $n$ in the node set $N$, we define a characteristic function or mapping $f: n \rightarrow c$, where $c\in C$ is a character set that classifies each node into exactly one category in $C$. The inverted list $\{I_i\}$, then, is defined as as 
$I_i = \{ n | f(n) = c_i \}, c_i \in C.$
There are two common types of inverted lists: content and clustering-based. 

\textbf{Content-based inverted index} characterizes nodes based on whether they contain specific content. In sparse vector search, such as WAND~\cite{wand}, vectors are indexed by their non-zero dimensions. As shown in Figure~\ref{fig:ivf}(a), each sparse vector typically contains only a few non-zero entries among dimensions $D_0$ to $D_n$. The index groups vectors by the presence of non-zero entries in dimension $D_i$. 
At query time, vectors with no overlapping dimensions with the query can be skipped, reducing the search space to roughly $O\left(\frac{||q||_0}{n+1}\right)
$, where $||q||_0$ is the $L_0$-norm (or the number of non-zero elements) of query $q$. However, since each sparse vector in the dataset typically has hundreds of non-zero values, the nodes involved in the search exceed the ideal number $|N| \times \frac{||q||_0}{n+1}$ hundreds of times, where $N$ is the number of vectors in the database. These methods require extra memory for intermediate results for computing distances between the query and all relevant vectors.

\textbf{Clustering-based inverted index} is widely used in dense vector ANNS, such as IVF~\cite{faiss} and ScaNN~\cite{scann}. This approach partitions the space by clustering techniques, like k-means, and groups nodes from the same cluster into a shared inverted list. As shown in Figure~\ref{fig:ivf}(b), the dataset is split into seven clusters, resulting in seven arrays in the inverted list. During ANNS, the query can skip clusters where the cluster centroids are too far from the query point, thereby narrowing the search space. The accuracy of this search depends on the number of clusters: larger numbers of clusters reduce time spent on centroid-query distance checks but increase the chance of false-negative cluster skips. However, for sparse vectors, which typically have significantly higher dimensions (tens to hundreds of thousands) and sparser distributions, using larger clusters poses challenges, resulting in either lower accuracy or minimal query speed improvements.

Inverted index acceleration has been explored using ASIC~\cite{ANNA}, NMP~\cite{MemANNS, REIS} and FPGA~\cite{fpga_anns, Chameleon}, but these designs target dense vector ANNS and perform poorly in sparse settings. First, they assume dense distance computation, which is inefficient for high-dimensional vectors with only a few non-zero entries (e.g., typically less than 5\% for the text embeddings~\cite{splade}). Second, they rely on clustering-based indexing, where each record belongs to one list, whereas in content-based indexing, records appear in multiple lists, requiring random access. Their batch-optimized access patterns break down under this access model.

In summary, content-based inverted indexing methods like WAND~\cite{wand} rely on sparsity and high-impact dimensions to prune effectively. However, SPLADE~\cite{splade} and similar models generate embeddings with many softly-weighted non-zero dimensions, which weakens WAND’s pruning power and increases candidate evaluations~\cite{splade_study}. In contrast, clustering-based inverted indexing struggles with high-dimensional sparse spaces, especially as the dataset size grows, leading to inefficiencies. 
Furthermore, profiling result with Intel Vtune shows IVF searches are highly memory-bound, with DRAM-bound rates exceeding 80\%. These challenges highlight the need for a hybrid indexing strategy that leverages both indexing paradigms and motivates exploring NMP techniques to alleviate memory bandwidth constraints.

\section{\design Overview}

To address these issues and enhance throughput for sparse vector ANNS, we propose \design, a hardware-software co-design leveraging NMP to efficiently utilize memory bandwidth and exploit inherent sparsity. \design introduces an NMP-friendly hybrid indexing algorithm, as illustrated in Figure~\ref{fig:ivf}-(c), combining content-based indexing’s strength in sparse data handling and clustering-based indexing’s scalability. 
The content-based index maps each non-zero dimension to the set of vectors containing that dimension, enabling early pruning of irrelevant data. Within each inverted entry, similar vectors are further clustered to support efficient lookup. A separate forward index stores the full original vectors, which the inverted entries reference to ensure accurate retrieval.
The detailed algorithmic optimizations for NMP-friendly processing behind this indexing scheme are discussed in Section~\ref{sec:alg}.
Similar to the vector database~\cite{milvus}, our method prioritizes inner product distance, a primary metric used in IR~\cite{sketch}.

SpANNS’s architecture is built to execute this hierarchical indexing strategy efficiently. It is implemented as a CXL Type-2 accelerator that interfaces directly with the host system. The architecture incorporates NMP-accelerated DIMMs and enhanced controller logic within the CXL device. Memory-intensive operations are offloaded to NMP kernels inside the DIMMs, while the controller manages orchestration, inter-DIMM coordination, and host communication. This design minimizes host-device data transfer and fully leverages NMP capabilities for high-throughput sparse ANNS.

\section{Hybrid inverted index for sparse vector ANNS}
\label{sec:alg}

A key challenge in sparse vector ANNS is that clustering-based inverted indices—commonly used in dense-vector systems like IVF~\cite{faiss}—become ineffective when applied directly to sparse data. This is because traditional centroids aggregate all non-zero dimensions across a cluster, producing dense vectors that poorly reflect the sparse structure of individual records. To address this, \design introduces a hybrid inverted index: content-based filtering is used first to narrow the search space to relevant dimensions, and within each inverted entry, vectors are grouped into clusters. Instead of centroids, we introduce \textit{silhouettes} to represent each cluster with the sparsity structure and the most representative dimensions. 

\begin{figure}[t] 
\centering
\includegraphics[width=\linewidth]{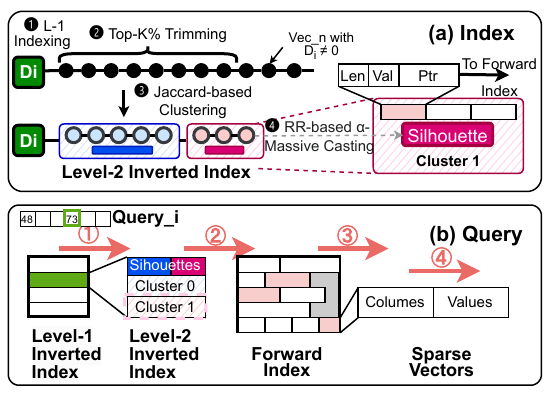} 
\vspace{-0.4cm}
\caption{Hybrid Inverted Index for Sparse Vector ANNS}
\vspace{-0.6cm}
\label{fig:overview}
\end{figure}

The process involves two main stages: building the two-level hybrid inverted index (Figure~\ref{fig:overview}-(a), \tikzcircle{1}-\tikzcircle{4}) and performing queries based on these pre-built indices (Figure~\ref{fig:overview}-(b), \tikzcirclewhite{1}-\tikzcirclewhite{4}). The hybrid inverted index in \design is built upon the conventional content-based inverted index (level 1) with an added clustering-based inverted index (level 2). \design also preserves the forwarded index to keep original data as most vector databases manage systems~\cite{milvus,pg}.

In the offline indexing phase (Figure~\ref{fig:overview}a), \tikzcircle{1} \design iterates through each record vector $x$. For every non-zero dimension  $D_i$ in $x$, $x$ is added to the corresponding inverted list: $I_{D_i} = \{ x\in X | x[D_i] \neq 0 \}$. \tikzcircle{2} Next, each record $x \in I_{D_i}$ is examined, and only those with top-K\% values for dimension $D_i$ are retained. 
This filtering step is inspired by WAND~\cite{wand}, as lower values of $x[D_i]$ contribute less to the inner product and are typically skipped during query-time early termination.
\tikzcircle{3}~Then, top-$K\%$~of non-zero values for each record are retained to reduce the set size of the union of non-zero dimensions in each cluster. To form these clusters, \design uses the Jaccard distance metric within k-means to optimize clustering for sparsity. \tikzcircle{4}~For each cluster, \design constructs a silhouette to summarize the sparse vectors. This process begins with an element-wise max operation for each non-zero dimension to produce a summary vector $m$, where $m[j] = \max_{x \in I_{D_i}} x[j]$. From $m$, the smallest subset $s$~is chosen to meet the constraint that its $L1$-norm preserves at least $\alpha$~of $m$’s $L1$-norm, following the $\alpha$-massive summary method from previous work~\cite{seismic}:
$$s = \arg \min_{s \subseteq m} ||s||{L_0}, \ \text{s.t.} \ ||s||{L_1} \geq \alpha ||m||_{L_1}.$$
However, to ensure fair representation of each record in the cluster, \design proposed the round-robin-based $\alpha$-massive method to build the silhouette. Specifically, we pick the largest non-zero dimensions in a round-robin manner across records, maintaining uniform overlap of non-zero dimensions between the silhouette and individual records.

During a query (Figure~\ref{fig:overview}b), \design performs the following steps in the proposed NMP accelerator:
\tikzcirclewhite{1} It probes the level-1 content-based inverted index, processing dimensions in order of their query values. This reflects the fact that the larger dimension tends to contribute more to the inner production result. 
\tikzcirclewhite{2} For each dimension, it compares the query to the silhouettes of associated clusters and prunes those unlikely to match if the query-silhouette distance is too far.
\tikzcirclewhite{3} For the remaining clusters, candidate records are fetched from the forward index. 
\tikzcirclewhite{4} Finally, the candidates are reranked by computing inner products with the query.

\begin{figure*}[t] 
\centering
\includegraphics[width=\linewidth]{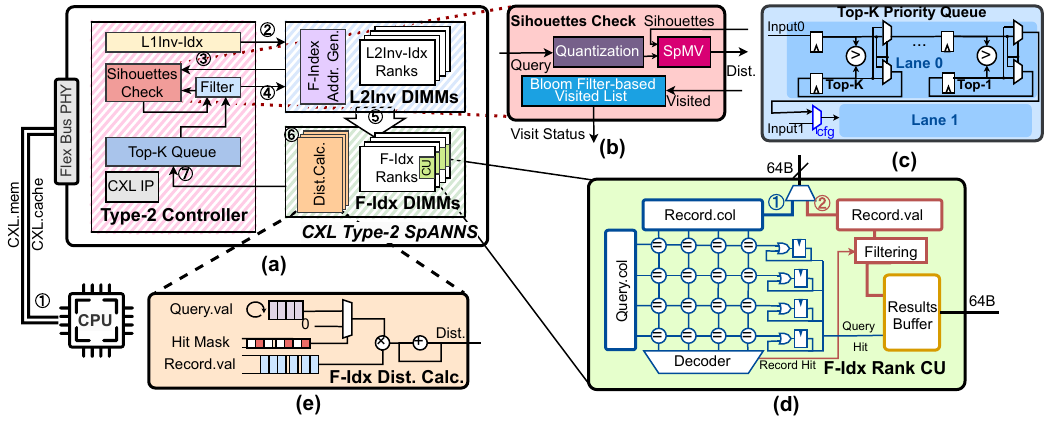} 
\vspace{-0.4cm}
\caption{Architecture of \design: (a) Dataflow and Components, (b) Silhouette Check Logic in Type-2 Controller, (c) Configurable top-K priority queue, (d) Compute Unit for Data Compact in Forward Index Ranks, and (e) Distance Calculation Module in Forward Index DIMMs.}
\vspace{-0.6cm}
\label{fig:arch}
\end{figure*}

To make the data structure NMP-friendly, we optimized both layout and access patterns to reduce memory traffic and improve locality. First, silhouettes from the same dimension are stored contiguously, enabling sequential access during query step \tikzcirclewhite{2}, which aligns with NMP’s bandwidth-efficient sequential data access pattern.  Second, we optimized the forward index layout to avoid cross-page access overhead. Each vector is placed entirely within a single memory page (DRAM page in our design), as long as it fits. This avoids fetching multiple pages for a single vector, which is especially costly in near-memory settings where memory page granularity dominates access efficiency.

\section{\design Near  memory accelerator}
\subsection{Architecture overview}
The Figure~\ref{fig:arch}-(a) shows the architecture of \design, a CXL (Compute Express Link) Type-2 accelerator tailored for efficient sparse ANNS. CXL Type-2 accelerators enable high-speed, in-memory processing and data sharing across hosts, minimizing data transfers between client CPUs and \design, and eliminating the need to duplicate indexed databases. 
The architecture consists of a Type-2 controller interfaced with specialized memory modules—Level-2 inverted (L2Inv) index DIMMs and forward index (F-Index) DIMMs—that work together to execute a two-level hybrid indexing process. Memory-intensive operations are distributed across these DIMMs to maximize efficiency and throughput.

The \textbf{Type-2 controller} orchestrates the query process by managing Level-1 index access, filtering candidates, and enabling direct inter-DIMM communication. It allows \textbf{L2Inv DIMMs} to forward candidate pointers directly to the \textbf{F-Index DIMMs}, bypassing the CPU and reducing data movement. The \textbf{F-Index DIMMs} store full records and include compute units to perform final distance calculations for final reranking.

\subsection{\design Dataflow}
Figure~\ref{fig:arch}-(a) shows the data flow in \design, proceeding through steps \tikzcirclewhite{1} to \tikzcirclewhite{7}. \tikzcirclewhite{1}
The CPU parses the query and sorts its non-zero dimensions by value. With a small number of non-zeros (typically 10–50 in queries), this step is lightweight and better suited for the host’s low-latency, high-frequency cores. Offloading it to the accelerator offers little gain. A software-managed buffer queues parsed queries to decouple preprocessing from NMP execution, enabling asynchronous, pipelined operation.
\tikzcirclewhite{2} The sorted query is then sent to the Type-2 controller, which probes the level-1 inverted index to identify relevant dimensions.  \tikzcirclewhite{3} Type-2 controller compares the query with cluster silhouettes stored in L2Inv DIMMs to determine which clusters are close enough to the query to justify further processing.  \tikzcirclewhite{4} Type-2 controller filters out clusters by comparing the distance between each cluster’s silhouette and the query. A top-K queue is maintained for the most promising records. The cluster is discarded if the distance exceeds a threshold (defined as $\beta$ times the current k-th best inner product in the top-K queue). For clusters that pass, any records already processed in the forward index are filtered out, ensuring only unique, relevant records are sent to the L2Inv DIMMs.
\tikzcirclewhite{5} L2Inv DIMMs send pointers of the selected candidate records directly to the F-Index DIMMs, reducing data movement and avoiding CPU. \tikzcirclewhite{6} The F-Index fetches the full records and computes refined inner products using shared non-zero dimensions. \tikzcirclewhite{7} Final scores are added to the top-K queue and returned to the host.

\subsection{\design Design Details}
\label{sec:design_detail}

The \textbf{Type-2 Controller} is responsible for accessing the Level-1 inverted index stored in a 1~MB buffer, shown in Figure~\ref{fig:arch}-(a). It can store up to 256K entries, which is typically sufficient for most applications. For instance, the BERT model’s vocabulary size is 30,522~\cite{bert}, which comfortably fits within this capacity. If the dataset dimensions exceed this 1~MB limit, the controller applies an LRU replacement strategy at the page level to manage memory efficiently. The detailed silhouette check is shown in Figure~\ref{fig:arch}-(b). 
To evaluate cluster relevance, the query is first quantized into 16-bit fixed-point format to reduce data size and enable efficient computation. A sparse matrix-vector multiplication (SpMV) unit then computes the distances between the quantized query and stored cluster silhouettes. We implement the SpMV unit following the architecture described in~\cite{ell_impl}. To avoid redundant checks, a Bloom filter–based visited list tracks previously processed clusters using a compact bit array and lightweight integer hash functions described in ~\cite{hash}, which are composed of XOR, shift, and add operations—all hardware-friendly. The computed distances are sent to the top-K priority queue (Figure~\ref{fig:arch}c), which dynamically updates the best candidates as the query progresses, setting a threshold (based on the k-th best distance) used in silhouette-based cluster filtering to exclude less relevant clusters. The top-K queue contains M parallel lanes, which can operate independently or be merged to support a larger Top-nK result. This flexible design allows \design to adapt to varying retrieval requirements.

\textbf{L2Inv DIMMs} store silhouettes at the start of each Level-2 inverted index entry using the Ellpack format~\cite{cusparse}. Silhouettes matrix is formed by stacking sparse silhouette vectors into a dense row-wise matrix, where most non-zeros are concentrated in a few columns. Ellpack format efficiently handles this structure by aligning non-zeros across rows, enabling compact storage and parallel access. The DIMMs also include an address generator implemented by lookup tables to translate candidate pointers and filter out distant clusters, forwarding only relevant candidates to the F-Idx DIMMs.

\textbf{F-Idx DIMMs} store complete forward index and perform final distance calculations. Within the F-Idx DIMMs, the rank-level compute unit (Figure~\ref{fig:arch}-(d)) consists of a comparator array (blue) and a compressed format filter (red). \tikzcirclewhite{1} The comparator array matches the query’s non-zero dimension indices against those of the record, identifying shared non-zero dimensions.
\tikzcirclewhite{2} The filter stage extracts the corresponding values of the matched dimensions from the record. Both the hit mask and filtered values are temporarily stored in a results buffer, decoupling DRAM access from the computation pipeline.
The distance calculation module (Figure~\ref{fig:arch}e) then performs the inner product computation. Leveraging the hit mask generated by the comparator array, the operation reduces to a multiply-accumulate (MAC) over only the matched non-zero dimension. The module completes in $||q||_0$ cycles. 
Each rank in DDR5 is composed of multiple dies (e.g., four dies for $\times$16 width configurations). To avoid inter-die communication, \design issues multiple burst reads, fetching 4$\times$64 bytes each time. This data layout ensures that each record’s column-value pair is stored within the same die, minimizing latency.
With DDR5 $\times$16, each DRAM die has a page size of 2~KB, allowing the forward index to be packed into 8~KB (2~KB $\times$ 4) bins (discussed in Section~\ref{sec:alg}) so that only one-row activation (ACT) command is needed per record.

\subsection{Query-aware Runtime Optimization}
Because the query's structure, such as the distribution of non-zero values and the L0-norm, cannot be determined during index time, \design proposed two query-aware runtime optimizations for more efficient processing.

\textbf{Dynamic forward index distance calculation:}  In the F-Idx DIMMs, as discussed in Section~\ref{sec:design_detail}, the latency of distance calculation is influenced by the number of non-zero values in the query. This can be further optimized based on the fact that the inner product distance computation between a sparse query and a record is symmetric: the operation yields the same result regardless of whether we iterate over the query or the record. Depending on whether $||r||_0$ or $||q||_0$ is smaller, it generates the hit mask from one vector and accesses values from the other, allowing the distance calculation to complete the computation in fewer cycles. This dual-mode flexibility enables \design to adapt efficiently to varying sparsity patterns at runtime.

\textbf{Load-balanced forward index checking:} Forward index checking can be distributed across different ranks and processed concurrently, with each rank determined by the candidate pointer. Strictly maintaining the order of F-Index checks, as dictated by the L2Inv index, allows the use of the current top-K distance to determine whether a cluster can be skipped based on the distance between the query and its silhouette. However, this strict ordering requires processing the current cluster completely before moving to the next, creating barriers that can stall processing and reduce the utilization of the F-Idx DIMMs.
To address this, \design relaxes strict processing order constraints using multiple delay queues, enabling out-of-order F-Idx record checking. This reduces stalls and increases F-Idx DIMM utilization, but may require checking more clusters as delayed processing lowers early termination accuracy based on the top-K distance. Balancing delay queues ensures efficiency while minimizing unnecessary cluster evaluations. Trade-offs are analyzed in Section~\ref{sec:load}.

\begin{figure*}[t]
\centering
\begin{minipage}[b]{0.32\textwidth}
  \centering
  \includegraphics[width=\linewidth]{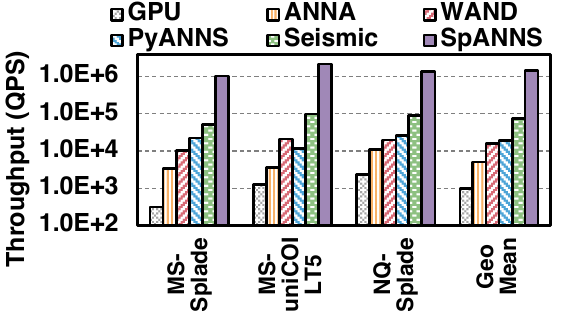}
  \captionof{figure}{Throughput of \design vs. GPU~\cite{cusparse}, ANNA~\cite{ANNA}, WAND~\cite{knowhere}, PyANNS~\cite{pyanns}, and Seismic~\cite{seismic}.}
  \label{fig:througput}
\end{minipage}
\hfill
\begin{minipage}[b]{0.32\textwidth}
  \centering
  \includegraphics[width=\linewidth]{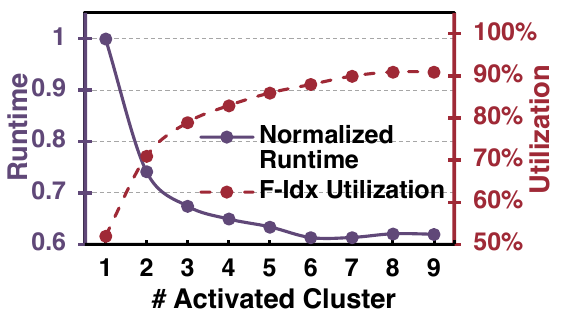}
  \captionof{figure}{Load-Balancing Trade-Off: Impact of Activated Clusters on Runtime and F-Idx DIMM Utilization.}
  \label{fig:load}
\end{minipage}
\hfill
\begin{minipage}[b]{0.32\textwidth}
  \centering
  \includegraphics[width=\linewidth]{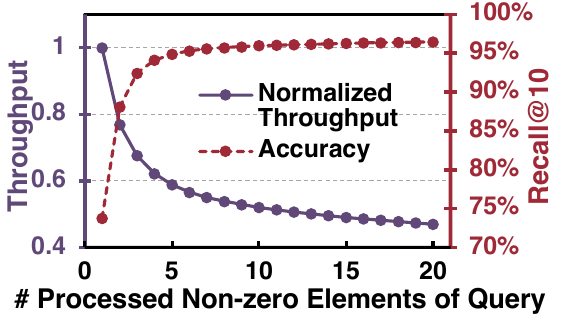}
  \captionof{figure}{Impact of Processed Non-Zero Query Elements on Throughput and Accuracy.}
  \label{fig:svr}
\end{minipage}
\vspace{-0.5cm}
\end{figure*}

\begin{table}[t]
\centering
\caption{Configuration Parameters for \design Evaluation}
\label{tab:config}
\begin{tabular}{|l|c|}
\hline
\textbf{Parameter}                & \textbf{Value}     \\ \hline
DRAM Configuration                & 8~Gb x16 DDR5-4800  \\ \hline
Timing (tRCD-tCAS-tRP)            & 34-34-34           \\ \hline
Channels / Ranks per Channel      & 8 / 8              \\ \hline
L1Inv-Idx Size                    & 256K entries       \\ \hline
L2Inv DIMMs : F-Idx DIMMs Ratio   & 1 : 3              \\ \hline
\end{tabular}
\end{table}

\section{Results}
\subsection{Experimental Setup}
\textbf{\design Setup:} We implemented a cycle-accurate simulator for \design based on Ramulator2~\cite{ramulator2} to evaluate performance, with configuration details provided in Table~\ref{tab:config}. The \design logic was developed in SystemVerilog, and timing and area metrics were evaluated using the Design Compiler with the ASAP 7~nm technology node~\cite{asap7}. The computing units integrated into the DRAM stack were synthesized at 300~MHz, matching the DDR5 internal frequency. To account for the lower transistor density in DRAM manufacturing compared to logic processes, we applied a $10\times$ scaling factor to the area overhead for both SRAM and compute units integrated into the DRAM dies~\cite{dram_factor}. The additional logic in the Type-2 controller was synthesized for 1~GHz. We modeled the area of the SRAM-based buffers using FinCACTI~\cite{fincacti}.

\textbf{Baselines:}  We evaluated \design against four baselines: (1) \textsf{GPU}: Due to the lack of GPU-based index structures for sparse ANNS (SANNS), we implement an exhaustive search baseline using the cuSPARSE~\cite{cusparse}. (2) \textsf{ANNA}~\cite{ANNA}: A clustering-based inverted index accelerator originally designed for dense vectors. Since no SANNS accelerator exists, we evaluate it using a dense-format conversion of the sparse dataset.
 (3) \textsf{WAND}~\cite{wand} algorithm for SANNS optimized in Knowhere\cite{knowhere}, the vector engine of vector database~\cite{milvus}, running on CPU; (4) \textsf{PyANNS}~\cite{pyanns}, a graph-based, HNSW-like SANNS on CPU; and (5) \textsf{Seismic}~\cite{seismic}, an inverted index-based SANNS on CPU. All CPU baselines were evaluated on an Intel Max 9460 with 80 threads and 512 GB DDR5 memory. The GPU baseline used an 80 GB NVIDIA A100.

\textbf{Datasets:}
We evaluated \design on three pre-embedded datasets~\cite{datasets}: the MS MARCO dataset~\cite{ms_marco} embedded with SPLADE~\cite{splade} and uniCOIL-T5~\cite{unicoil}, denoted as \textsf{MS-Splade} and \textsf{MS-uniCOILT5} (8.8M records, 7K queries), and the Natural Questions from the BEIR~\cite{NQ}, embedded with SPLADE~\cite{splade}, denoted as \textsf{NQ-Splade} (2.7M records, 8K queries).

\subsection{\design Performance}

Figure~\ref{fig:througput} presents the throughput of \design compared to the baselines across three datasets. We performed a grid search to optimize the parameters of each algorithm for the best throughput while maintaining $\text{Recall}@10>0.9$. \design is $15$–$22\times$, $46$–$180\times$, and $69$–$101\times$ faster compared to the CPU-based indices Seismic~\cite{seismic}, PyANNS~\cite{pyanns}, and WAND~\cite{knowhere}, respectively. Compared to exhaustive forward index computation on a GPU, \design is on average $1500\times$ faster. Although GPUs excel at accelerating sparse matrix computations, naive exhaustive search for the forward index is inefficient, resulting in such a large speedup of \design. Even the slowest CPU index in our experiments (WAND~\cite{knowhere}) is up to $32.5\times$ faster than the GPU. To the best of our knowledge, no GPU libraries currently exist to accelerate sparse vector ANNS using index-based methods, such as those in vector database search engines. 

\design also outperforms the clustering-based inverted index accelerator ANNA~\cite{ANNA} by up to 280$\times$. This speedup stems from two key factors: (a) \design supports optimized sparse-vector distance computation, avoiding the overhead of dense arithmetic in ANNA; and (b) \design's hybrid index design significantly reduces the number of distance calculations needed compared to cluster-only indexing.

The inverted index methods (\design, WAND~\cite{wand}, and Seismic~\cite{seismic}) tend to outperform the graph-based method (PyANNS~\cite{pyanns}) when the embedding model generates more sparse embeddings. For example, embeddings generated by uniCOIL-T5 have $1.8\times$ fewer non-zero values compared to those generated by SPLADE on the MS MARCO dataset. \design is $180\times$ faster vs. PyANNS~\cite{pyanns} for MS-uniCOIL-T5, compared to a $46\times$ for MS-SPLADE.

\subsection{Load Balancing Analysis}
\label{sec:load}

Strictly maintaining the order of F-Index checks (single activated cluster) leads to low F-Idx DIMM utilization, as barriers from strict ordering stall ranks waiting for previous clusters to finish. 
Strictly enforcing the order of F-Index checks, where only one cluster is processed at a time, results in low F-Idx DIMM utilization. This strict ordering creates barriers that cause processing units in other ranks to idle while waiting for the current cluster to finish.
Figure~\ref{fig:load} shows the effect of relaxing these constraints on F-Idx DIMM utilization and runtime. With one activated cluster, about 50\% of F-Idx DIMM processing units and bandwidth remain idle. Relaxing the strict processing order of clusters increases utilization but comes at the cost of additional cluster evaluations. Activating more clusters simultaneously reduces the efficiency of the L2Inv index’s filtering, leading to unnecessary processing of less relevant clusters. However, the benefit of increased utilization outweighs the overhead of additional cluster checks when the number of activated clusters is fewer than five.

Based on this analysis, \design activates five clusters simultaneously in its optimal configuration. At this level, F-Idx DIMM utilization reaches around 90\%, and runtime is reduced significantly. We observed that the accuracy impact from processing slightly more clusters is negligible, with improvements in recall being less than 0.2\%.

\subsection{Early Termination for Query Optimization}

Figure~\ref{fig:svr} shows that \design can terminate early by processing only a subset of query elements. \design processes query dimensions in descending order of their values, as higher-valued dimensions contribute more to the inner product. Figure~\ref{fig:svr} shows it reaches 95\% $\text{Recall}@10$ after just the top five dimensions. Beyond that, throughput drops by nearly 20\% with little gain in accuracy. This early termination strategy, combined with final reranking, allows \design to reduce computation while maintaining high accuracy.

\subsection{Area and Power Overhead Analysis}
\begin{table}[t]
    \centering
    \setlength{\tabcolsep}{2pt} 
    \vspace{-0.0cm}
    \caption{\design Overhead}
    \begin{tabular}{lcc}
    \toprule
    \textbf{Component} & \textbf{Area ($\text{mm}^2$)} & \textbf{Power (W)} \\ \midrule
    Type-2 Controller  & 0.818                        & 1.22               \\
    L2Inv DIMM         & 0.117                        & 0.015              \\
    F-Idx DIMM         & 0.503                        & 0.127              \\ \bottomrule
    \end{tabular}
    \label{tab:area}
    \vspace{-0.5cm}
\end{table}

Table~\ref{tab:area} shows the area and power consumption of \design components. The proposed design adds only 0.82 mm² of area and 1.22 W of power as a CXL Type-2 device—significantly smaller than commercial Type-2 accelerators equipped with 16 ARM cores~\cite{commercial_type2}, where each core alone occupies approximately 2.5mm²~\cite{arm_size}.  The additional power for the DIMMs is 0.015~W and 0.13~W per DIMM, which is less than 2\% of the typical 10~W power consumption per DIMM, ensuring minimal impact on the overall system. \design indices can be built on the CPU within 15 minutes for all three datasets used in our experiment, compared to 4 hours to build the PyANNS~\cite{pyanns} index. The major reason is that our design computes clustering only on the trimmed L1 index list, whereas PyANNS must search across the entire dataset.

\section{Conclusion}
\design bridges the gap in sparse vector search acceleration, achieving significant improvements in performance over CPU and comparable scalability to dense accelerators. Its hybrid inverted index and near-memory architecture enable efficient sparse ANNS for vector search engines. The architecture offloads memory-intensive tasks to NMP-enabled DIMMs and minimizes host-device data movement through direct inter-DIMM coordination. 

\section{Acknowledgment}
This work was supported in part by PRISM and CoCoSys—centers in JUMP 2.0, an SRC program sponsored by DARPA—and by the National Science Foundation under Grants No. 2112665, 2112167, 2003279, 2120019, 1911095, 2052809 and 2211386.


\bibliographystyle{IEEEtran}
\bibliography{ref}

\end{document}